# Cryptographic Key Management for Smart Power Grids

## *Approaches and Issues*

M. Nabeel, J. Zage, S. Kerr, E. Bertino
*CS Department, CERIAS and Cyber Center, Purdue University*

N. Athula. Kulatunga, U. Sudheera Navaratne
*ECET Department, Purdue University*

M. Duren
*Sypris Electronics*
February 22, 2012

**Abstract**

The smart power grid promises to improve efficiency and reliability of power delivery. This report introduces the logical components, associated technologies, security protocols, and network designs of the system. Undermining the potential benefits are security threats, and those threats related to cyber security are described in this report. Concentrating on the design of the smart meter and its communication links, this report describes the ZigBee technology and implementation, and the communication between the smart meter and the collector node, with emphasis on security attributes. It was observed that many of the secure features are based on keys that must be maintained; therefore, secure key management techniques become the basis to securing the entire grid. The descriptions of current key management techniques are delineated, highlighting their weaknesses. Finally some initial research directions are outlined.

## 1. Introduction

Smart meters provide an economical way of measuring energy consumption that allows utility companies to introduce different prices based on the time of day and the season. For the consumers, a smart meter can display current information on consumption and can update time of use tariff changes in in-home displays allowing consumers to better monitor and manage their energy usage. Smart meters and corresponding infrastructure provides more information for smart appliance manufactures to build efficient and effective products. More progressive applications are in tele-health and social care services that may reduce the burden on government health services and enable longer independent living. Vulnerable and low income consumers can also be targeted for assistance. Smart meter technology can provide both social and economic benefits. To accrue these benefits, each meter must be able to reliably and securely communicate the information collected to a central location. Considering the varying environments and locations where meters are found, such problem will not be easily solved [24].

In a recent Scientific American blog, the question "Could Hackers Break into Your Electric Meter?" was debated. The answer was a definite yes [17]. Many utility consumers are suspicious or anxious about a utility remotely reading a meter, shutting off power to a house or modulating individual appliances to shed load during peak hours. Consumers may be doubly concerned that these same actions can also be performed by hackers. Other researchers and security firms have demonstrated these vulnerabilities [3][13][14]. When power meters came into existence, providers had to deal with attempted frauds. Today, there are published web pages dedicated to out-smarting the manual meters. The smart meter still possesses many of the previous problems of its manual counterpart plus the additional vulnerabilities created by the computer and communication technology. For example, a software patch that underreports usage is much harder to detect than previous physical alterations. The utility can spot an altered meter, but a change in software is physically undetectable. Currently, electricity theft accounts for one to two percent of U.S electricity production, but thefts may increase because of the ease of tampering and the invisibility of the action. However, theft may not be the only major concern. As hundreds of millions of smart meters and devices eventually become connected to the power grid, new risks will be introduced into the system. The focus now changes from theft of service to all kinds of mischief. Basically, with every new device connecting to the grid, the attack surface grows and opportunities for damage increase. One goal of smart meters is the ability to control device. What if hackers caused vast numbers of appliances to turn on and off at inopportune moments? The grid can face a considerable amount of instability depending on the implementation (level of control and monitoring supported in substation level). Interfering with monitoring and data collections exposes the system to all types of vulnerabilities, including denial-of-service attacks. Moreover, as the grid becomes more automated, threats will move more swiftly [30].

In late August 2011, nCircle, a provider of automated security and compliance auditing solutions, announced the results of their 2011 Smart Grid Survey taken by 544 respondents from the IT security industry. When asked in the survey "Are you concerned about smart grid cyber security?" 77% responded "yes". When the surveyors were asked to comment about this they replied: "It's not surprising that the majority of respondents are concerned. The smart grid initiative involves aggressive deployment of a network device – in this case a smart meter – to nearly every household in America. That's quite a target surface for a Stuxnet-type attack"[27].The Stuxnet attack proved that utilities' seemingly isolated networks could be compromised, potentially disrupting energy production and distribution.

Securing the communications in a smart-meter infrastructure, which we refer in what follows as Advanced Metering Infrastructure (AMI), requires combining different security technologies. Among these technologies, encryption represents a crucial technology, as it allows one to secure the data transmitted across the AMI and authenticate the different parties involved in interactions across AMI. However, even though many different encryption technologies exist, their application to AMI poses several challenges, one of which is scalability, as an AMI easily involves millions of devices. Also as an AMI uses many different communication technologies, each potentially using different encryption protocols and requiring different keys, managing keys becomes a complex task.

In this report, we present an overview of approaches for managing encryption keys in AMI and main security issues. In what follows, we first present a high-level description of the organization of the AMI as this is crucial in order to understand the main requirements for key management. We then present an overview of communication protocols and standards adopted or proposed for AMIs, followed by a discussion about security of smart meters and networks, including ZigBee. We then discuss security threats and finally outline some initial research ideas. This analysis shows that a reoccurring and prominent source of the problems is key management, i.e., protecting and controlling access to the keys that underpin the encryption process.Descriptions of key management schemes tangential are then presented, followed by a short overview of our proposed key management scheme is an improvement over the available schemes.

## 2. Advanced Metering Infrastructure

An abstraction of an Advanced Metering Infrastructure (AMI) could enumerate four main components: the utility company (utility, for short), the data collector or concentrator often located in the neighborhood, the smart meter, and the home or office as shown in Figure 1. Other AMI topologies are possible, resulting from merging or duplicating these logical components. The AMI or the intelligent AMI "smart meter" is a combination of equipment, communications and processes for utilities to enhance their internal operations and customer service. Beginning with the smart meter, Figure 1 box 3 includes the smart meter board used to measure energy consumption and interface cards to communicate with the home area network (HAN) and the concentrator. The meter may also contain a disconnection function that allows one, if enabled, to remotely connect or disconnect the service. Smart meters form a collection of communicating devices within the same approximate geographic location that support the transportation of data from the individual meters to the WAN collector. This two-way wireless or wired communications network is labeled the Neighborhood Area Network (NAN).The home, box 4, provides the consumer access points to control and monitor consumption through remote control interfaces. It contains the home gateway to communicate with the meter which also supplies a path for the home display and controllers to the energy consuming devices. The first two boxes (1 and 2) in Figure 1 introduce additional capabilities over previous metering systems and are the most important to secure. The concentrator normally has two card interfaces, one to the meter (NAN interface) and one to the central office (WAN Interface). The concentrator may possess additional message processing capabilities and, depending on its purpose, may have a small amount of buffer memory to use between the two communication cards or larger amounts of memory to serve as a storage location. The central office connects to the concentrator through the WAN interface, interfaces with the event manager to process alarms and alerts. The Meter Data Management Service module in the utility is the processing unit for billing and system operations. Other modules, such as portals allowing customers to view meter data and bills,also can be present.

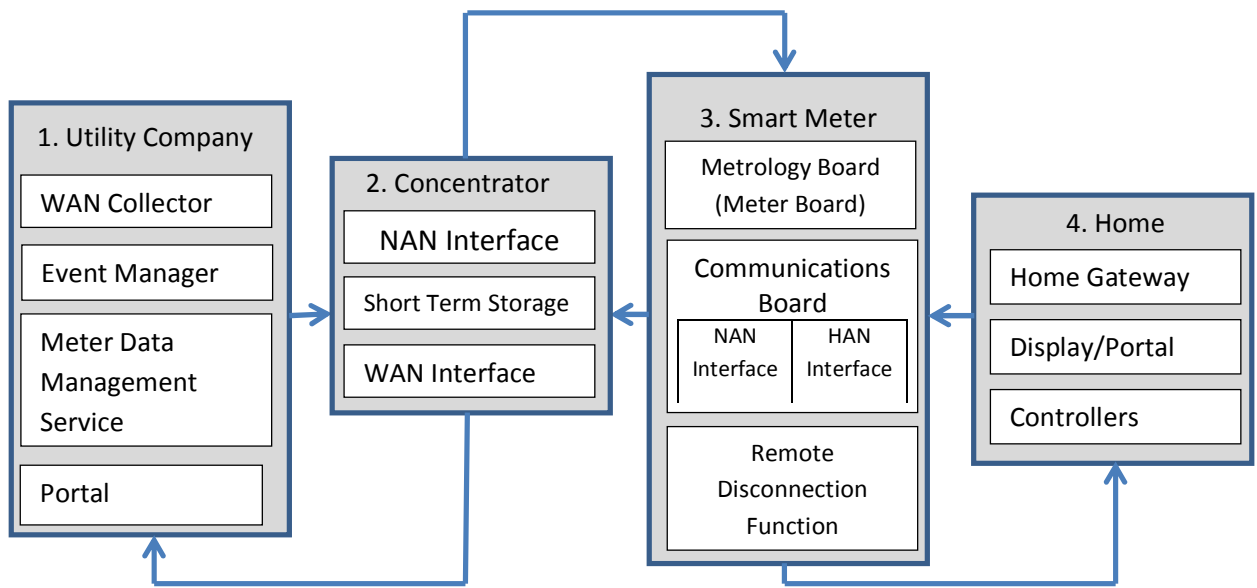

**Figure 1:**The Advanced Metering Infrastructure (AMI) consisting of four main logical components:1.the Utility Company, 2. the Concentrator, 3. he Smart Meter, and 4. the Home

Smart Metering, or an AMI, relies on a complex communications infrastructure to exchange information with and control digital meters and other devices that make up the AMI. This communications infrastructure can consist of various communications networks and requires frequent transmittal of measurements and command and control data. Typically, there are three or more networks, namely the HAN, the NAN, and the WAN. As it can be seen by the arrows in Figure 1, there are six communication paths in the AMI. These paths can vary from simple low speed power line carriers to very high bandwidth wireless systems. No one communication technology provides the entire coverage.

## 3. Communication Standards and Protocols

The connection between the smart meter and home appliances, that is, the HAN, can use several communication technologies such as *ZigBee, Wi-Fi, Ethernet, Z-Wave, HomePlug, Wireless M-Bus,* and *Wavenis*. This report will concentrate on the ZigBee technology. As in the HAN, the connection between the collector and the smart meter also can be implemented using different network technologies. Most commonly used are the RF Mesh and 3G (EDGE or HSDPA) networks because of the extensive coverage these typically offer in all environments. The connection between the utility and the collector also allows for many protocols and can use a GPRS-based protocol. Table 1 maps the communication technology and protocol standards for the various networks of the AMI.

**Table 1:** AMI network technology and protocols standards

| Wide Area Network (WAN) public/private | | | | | |
|---|---|---|---|---|---|
| substation | CoreMetro Network | | Backhaul Network | | |
| | wireline | wireless | wireline | wireless | |
| DNP3/IEC 61850 | IP/MPLS, SONET/STS-MESH/DWDM, PacketMetro-Ethernet | Wimax 802.16d/e, Trunked Radio Sidebands | GPON/EPON, RFoG-DOCSIS, Metro-Ethernet, DSL/POTS/PDH | 3G-3GPP/1X/RTT/EVDO/EDGE/HSDPA, Wimax 802.16d/e, Mesh RF/mm-Wave, RF Pto-to-PtoMAS, 802.16/LMDS | |

| Neighborhood Area Networks (NAN) public/private | | | | |
|---|---|---|---|---|
| substation | wireless | wireline | wireless | powerline |
| DNP3/IEC 61850 | Wimax 802.16d/e | RFoG-DOCSIS, FTTP/FTTH/Ethernet | 3G-3GPP/1X/RTT/EVDO/EDGE/HSDPA, RF Mesh, RF Radio Pto-Mpt/MAS, wLAN 802.11 n/g, 802.2.15.4/Zigbee | power line carrier(PLC), Broadband over powerline (BPL) |

| Home Area Network, Building Area Network, Industrial Area Network | | | |
|---|---|---|---|
| Smart Meters | | wireless | powerline |
| ANSI C.12.22 | | 802.11, Z-Wave, 6LoPan, 802.2.15.4/Zigbee | HomePlug |

Each network is a unique distributed computing environment that also may have a sub-network embedded to meet special requirements. One objective of the AMI is to provide the capability to enable an application in one particular network to communicate with an application in another network. One of the protocol standards that address each of the network domains (See Table 2), the North American Electric Reliability Corporation Critical Infrastructure Protection Standards (NERC CIP), clearly specifies that utilities must integrate information system security into all aspects of their automation systems. This includes not just devices, computer systems and technologies, but also policies, procedures, and training. As suggested by the NERC CIP, security must be integrated throughout the entire AMI. Tremendous amounts of data will pass through the AMI domains and often with very different security requirements (e.g., the sensitive metered data versus the ambient air temperature at the customer's site). No one security protocol can handle all these different requirements.

Also, a sub-group of the NIST Smart Grid Interoperability Panel-Cyber Security Working Group (SGIP-CSWG)has identified a number of *evident and specific security problems* that are not obviously solved by existing standards, de facto standards, or best practices [19]. Specifically, lacking are high-level designs for key protocols and interfaces resulting from the many issues encompassing the authentication and authorization users to various devices of the networks. An AMI authentication system must operate in a massively distributed and locally autonomous setting. To address this environment, the CSWG suggests secure authentication methods allowing for local autonomy when needed and providing revocation and attribution from a central authority as required. To establish a balance between central and local authority, a series of new standards that provides for a hybrid approach should be developed.

**Table 2:** AMI communications security protocol standards

| Networks | NERC CIP | IP-Sec | VPN Tunnel | Private APN | SSL | WS Security | LDAP | Zigbee ECC |
|---|---|---|---|---|---|---|---|---|
| | | | Security Standards | | | | | |
| Wide Area Network | ■ | ■ | ■ | backhaul | Enterprise | Enterprise | | |
| Neighborhood Area Network | ■ | ■ | ■ | | | | | |
| Home Area Network | | ■ | ■ | | | | | ■ |

As observed from the various choices among technologies listed in Table 1 and the security protocols/standards listed in Table 2, AMI is a complex and highly integrated network of multiple systems and technologies designed to measure, collect and analyze energy usage.

**ANSI Meter Standards**
Originally, the data formats, data structure and protocols for electricity meters were proprietary. However, utility companies wanted a compatible communication protocol between meters so they would not be restricted to a single vender. Thus, ANSI C12.19 was created to describe meter data formats and structures. Other additional protocols were defined such as ANSI C12.18 to provide point-to-point communication over optical connections and C12.21 for communication over telephone modems for meters. Users also wanted to send and receive C12.19 tables remotely over networked connections and ANSI C12.22 was inaugurated.

## 4. Smart Meter Design and Security

The NISTIR 7628 outlines security requirements for an AMI and defines the need to provide privacy and integrity of data exchanged between various components that make a metering system. A fully deployed AMI contains many working parts and communications paths and the priority (low, medium, high) of a given security principal (Confidentiality, Integrity, and Availability) varies widely from component to component and function to function. For Smart Meters, understanding the physical layout of the smart meter and the communicating networks is critical. The smart meter has two inner physical components, the meter board and the communication board. The meter board contains storage tables to hold the keys used for communication. Within the communication board are the communication protocols. The communication board and meter board are connected through a serial port. The optical port in the meter allows reads and writes to the C12.19 tables in the meter board. As in the C12.18 and C12.19 standards, access to the meter through the optical port is restricted by six security levels, L0 to L5, with the highest privilege being L5.

Critical to the security of the AMI is the design of the smart meter. How the smart meter secures its keys, how software updates are executed, and how the smart meter is managed can inform us of the effectiveness of the standards for communication between the smart meter and the collectors. In the current meter design, the answers are inadequate and open to security risks.

We now look at ANSI C12 standards for the smart meter security in detail. As stated previously, the meter board contains the physical storage for the keys and passwords. All data stored in the meter board are stored in cleartext. These values travel from the meter board to the communication board through a serial port in cleartext. The optical ports on the smart meter permit the operations of read and write to the tables in the meter board. In order to protect passwords and keys, the protected tables permit write only and return empty values when read.

Operators must log into the smart meter through the optical port or the wireless link to execute commands on the smart meter. To successfully execute the command, an operator's security level (L0 to L5) must be greater than or equal to the security level of that command. Access to each security level is protected by a password. However, the default password is easy to guess.

The base level of security is L0. L0 requires no password and only allows read commands configured at the L0 security level. L1 allows only the meter to be read, while L2 allows for demand resets. L3 permits entry of a series of commands for the meter maintenance, such as setting the date and time, modifying the Time-Of-Use rates and calendar, and loading profile configurations. L4 permits programming and procedures to be done except for device configuration. Finally, the highest level, L5 allows device configuration, which is needed when altering the ANSI standard tables or manufacturing tables within the meter.

Within the meter board of the smart meter are located the tables for meter data storage. The tables hold keys, passwords, and access permissions used in the smart meter protocols. There are six main tables used in relation to the security of the smart meter. The first is the Security/Password table, with a table id of 42. This table contains 5 passwords, one for each of the security levels L1 through L5. The next table, with a table id of 43, is the Default Access Control table. It specifies the default read and write permissions. The third table, the Access Control table, with a table id of 44, contains the access permissions for specific tables or procedures. Table 45, the Keys table, holds the keys used for encryption and authentication. Table 6, the Extended Keys table, holds application-related communication keys, an extension available on top of the smart meter security. The last table, table 46, called the Host Access Security table, is used to store authentication keys, encryption keys, and access permissions used by remote nodes.

There are inherent problems with the meter's design leading to the likelihood of compromising its integrity. The meter has very limited RAM which also limits the amount of program code and leaves less room for error checking. The device is susceptible to typical software flaws such as buffer overflows and state machine flaws [3]. It also has several hardware weaknesses which may lead to attacks such as bus sniffing attacks, clock speed and power glitching attacks [26]. Through varying timing and voltage levels, the attacker can create abnormal operating conditions thereby exposing access to secure, otherwise unnoticed parts of the system. Other invasive techniques can be used to generate faults, such as exposing the chip surface to laser light. Microprobing to inject signals, capture data, manipulate registers and thereby retrieve sensitive information is another method of accessing the system. Smarter techniques, such as differential power analysis, are used to extract the secret keys and circumvent security embedded in integrated circuits.

Key storage can also be a problem since many systems may store multiple copies of the secret keys in various locations possibly accessible to hackers. In a smart meter, it is reasonable to expect the presence of stored passwords and related encryption material due to the requirements of authentication and integrity protection on the meter. One set of possible keys stored are asymmetric keys used to authenticate the meter when it joins the Neighborhood Area Network (NAN). By obtaining this key, the attacker can impersonate the victim meter. Symmetric keys used for shared authentication also could be stored and then the attacker would have access to the other devices sharing that key. It is known that C12.18andthe C12.22 protocol leverage a shared secret for authentication [3].The meter itself is easily accessible and an attacker can extract and review the firmware and EEPROM data contents for keys. Tampering can also be carried out by service personnel; about 25% of all reported data security breaches are the result of insiders [26]. During meter maintenance, the service person can have access to secure data.

It is also important to recognize that there are multiple vendors of AMI equipment and due to the rapid development and changing implementation specifications, each vendor's implementation decision will greatly influence the security of the system [3]. For example, some vendors encrypt all traffic transmitted to the NAN while it is believed that others do not or leave it as a configuration option.

## 5. Network Design and Security

In AMI deployments there can be thousands of smart meters, one for each residence. The data from each meter must be delivered to a concentration point for further processing. Two basic categories of smart meter networks defined by the technology are Radio Frequency (RF) and power line carrier (PLC). RF wireless networks are commonly used in the current deployments because of their cost-effectiveness. The smart meters talk to each other (hop) to form a mesh of interacting nodes called a mesh network. The network topology can be full or partial. In the full mesh topology, each meter would be connected directly to each of the other meters within its network. In the partial mesh topology, some meters are connected to all the meters, but some of the meters are connected only to those meters with which they exchange the most data. The mesh networks with wireless devices provide self-adapting, multi-path, multi-hop communications between meter nodes. Mesh networks are considered very reliable because they provide redundant communication paths that compensate for failures from natural causes. In rural areas where distance or terrain poses challenges, the utility can use PLC especially if an existing utility infrastructure of poles and wires is available. Point-to-point RF technology is also used for smart meters sending information to a concentrator or collector. The collector transmits the data using various WAN methods to the utility central location.

Figure 2 shows a block diagram of the typical network connections that can be traced from the home to a concentrator node and then to the utility control center. The HAN devices are connected to a smart meter network using a mesh RF technology such as ZigBee. Also using ZigBee/mesh wireless the smart meters are aggregated into a mesh configuration for back haul to the utility control center. Each concentrator defines a sub-network NAN supporting communications to and from the NAN head-end. For those nodes, the concentrator establishes network time synchronization and coordinates overall operation of the wireless mesh network. Additionally, the concentrator provides local data storage as a form of data redundancy.

Concentrator nodes communicate with the utility through common communication mechanisms including the Internet. The system enables a two-way distribution of information between consumers, service providers and utility companies. As more AMI systems are deployed, inherent security risks dramatically increase. The system's reliability, consumer privacy, network integrity, interconnectivity of all participants, and the integration with critical utility applications are far more vulnerable than with traditional metering systems.

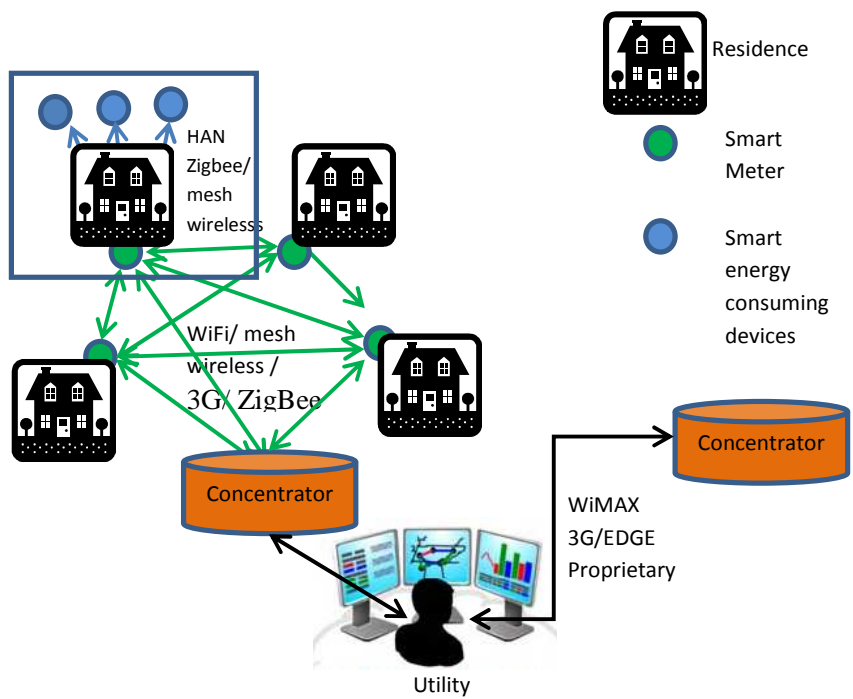

**Figure 2:** A block diagram of typical AMI components and connections

**ZigBee and ZigBee Profiles**
The Smart Energy market is turning to ZigBee for low-power wireless technology. To establish the necessary communications, two ZigBee networks may be required. One is the network of neighborhood meters for sub-metering within a home or apartment to transfer data from the electricity meters to base stations. The other is to possibly communicate with the devices in the office or home. ZigBee has other profiles intended for use in health care, home automation and commercial building automation. Different installations and preferences result in different network topologies and operations. However, each of these networks operates by applying the same basic principles to ensure interoperability. ZigBee is rated low-power, implemented using only 120 KB of memory and capable of operating on battery power for years.ZigBeecan be usedin applications that do not require high data rates. It can support data rates at maximum throughput of 250 Kbps and connects to lightweight embedded technology. The simplicity of the ZigBee protocol allows for system-on-a-chip implementations.

ZigBee, initially released in 2004, has undergone important revisions. In 2006, support for encryption and frame authenticity was added. A "trust center", a new security model, was created in 2007 and more security features were added to ZigBee-PRO. The new implementations have advanced networking and security features such as:
- Retries and acknowledgements.
- DSSS (Direct Sequence Spread Spectrum).
- Each direct sequence channel has over 65,000 unique network addresses available.
- Point-to-point, point-to-multipoint and peer-to-peer topologies supported.
- Self-routing, self-healing and fault-tolerant mesh networking[31].

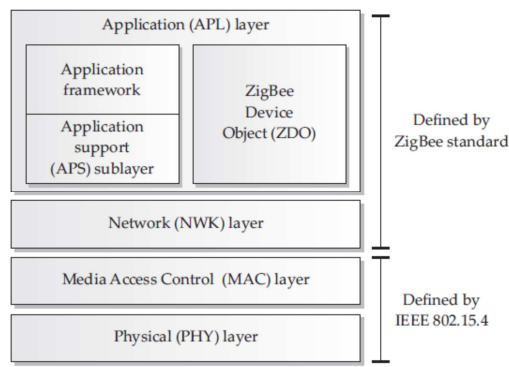

**Figure 3:** ZigBee Protocol Stack [34]

Figure 3 outlines the architecture of the ZigBee Protocol stack. The physical and MAC layers are defined by the IEEE 802.15.4 specification [10]. ZigBee allows a full mesh network, as described above. The ZigBee specification [2] builds upon these layers, defining the network layer and application layer, the ZigBee device object (ZDO's), and manufacturer-defined application objects which allow for customization. The routing and mesh capability are defined within the network layer. The ZigBee standard calls for AES 128-bit encryption with authentication and encryption at the MAC, NWK and Application levels. The security features can be implemented in any of the layers and can be defined specifically for the application framework by the profile.

**ZigBee Security**
ZigBee uses a hierarchy of keys to manage security. There are three types of keys: Master, Network and Link. The keys are distributed from a ZigBee "Trust Center". The Trust Center is a device trusted by all devices within a ZigBee network to distribute keys for the purpose of network and end-to-end application configuration management. There is only one Trust Center in a ZigBee network. The AMI has the trust center located inside the smart meter, where keys can be installed or generated.

Keys can be installed by three methods: out-of-band, in-band, factory pre-loaded. The out-of-band method loads the keys through mechanisms other than through normal network operational channels e.g., through a serial port on the ZigBee device attached to a laptop. The in-band method delivers the key through a normal communication channel. There is a small window of vulnerability where the not-configured device is unprotected. Finally, the key can be set at the factory. The key is then conveyed to the customer, hopefully in a secure manner. Also, in this case the vendor has knowledge of the customer's key. From a security standpoint, customer generated out-of-band loading of the key would be the most secure method. The Master Key, the address of Trust Center and the Trust Center keys can be pre-installed.

The Network Key is used to perform Network Layer security (routing messages, network join requests, etc.) and to prevent the unauthorized joining and use of the ZigBee network. All devices on a ZigBee network share the same Network Key. When one or more devices are compromised and the Network Key is leaked, there is a window of vulnerability until a new Network Key is established. A set of Network Keys is held by the Trust Center and the current Network Key is identified by a key sequence number. The Trust Center may periodically update and switch to a new Network Key. The new key is encrypted with the old Network Key and broadcasted. Devices are sent messages to switch to the new key. The Trust Center's Network Keys can be pre-installed. Each smart meter can also have a Network Key, and a Master Key pre-installed or in-band loaded.

Link Keys are secret session keys used between two communicating ZigBee devices and are unique to those devices. Link Keys are for unicast communication while Network Keys are for broadcast communication. Establishment of Link Keys is based on a Master Key which controls Link Key correspondence. Link Keys are used to secure unicast messages between two devices at the Application Layer. The Master Key is not used to encrypt frames; it is an initial shared secret between two devices performing the key establishment procedure to generate the Link Keys. The Master Key could be preconfigured or provided *in the clear* from the Trust Center. By possessing a correct Network Key, a node can utilize the network for receiving and transporting data. Notice that the Network Key provides a security check for utilizing the network. To provide source node authentication, an additional security protection, the Link Key(end-to-end crypto key) must be generated and used. This key is unique to a pair of devices that are communicating with each other and is derived from their respective Master Keys.

There are three modes to establish a key: the Symmetric Key Key Establishment (SKKE), the Public Key Key Establishment (PKKE), and the Certificate-Based Key Establishment (CBKE). The SKKE produces the link key based on a shared secret (Master Key). If the Master Key is compromised, publicly known, or set to the default value, the established link key is also compromised. The PKKE is based on a shared static or temporary public key. A public key does not need to be kept secret, so its knowledge does not compromise the link key. Depending on the method the public key was bound to a device, either transported independently or transported as part of authentication certificate, it is either PKKE or CBKE. The CBKE provides the most comprehensive form of key establishment [9]. It is based on the use of the Elliptic Curve Qu-Vanstone (ECQV) [38] implicit certificates that are much smaller in size than conventional certificates. Each such certificate binds a device MAC address and manufacturing identifier to an ECC key pair. The CBKE works similar to Diffie-Hellman key exchange protocol except that the public keys of the two devices are not exchanged but instead their corresponding certificates are used to derive the public keys of each other.

ZigBee provides two modes of operation: residential mode (Figure 4) and commercial mode (Figure 5). These modes affect the distribution (if any), storage, and encryption of the keys. In the residential mode, the list of devices, Master Keys and Network Keys can be maintained by the Trust Center or by the devices themselves. The Trust Center's responsibility in the residential mode is to maintain a Network Key and to act as the central authority for admitting nodes to the network. Application security is provided through a single Network Key. Underlying the frame security service is the IEEE 802.15.4 protocol. Each mode can identify what is encrypted and the key length. Frame counters given by the source to protect from a replay threat can also be established. In addition a Key Identifier field can specify the key required to communicate with a node.

In the commercial mode, the Trust Center maintains keys (Master, Network and Link) for every network device, allowing for centralized key control and maintenance. In this mode, the memory cost increases based on network size. The Master Key is a secret key between two nodes and is the starting point of establishing the Link Key. The Link Key establishment is processed by the SKKE protocol. The Link Key is shared by two devices and the Network Key is shared among all the devices. In the commercial mode, to transport the Network Key an additional key derived from a Link Key is used to protect Key Transport Messages carrying a key other than a Master Key (Key-Transport Service). The difference between the residential and commercial Network Keys is in the rules for distributing Network Keys to devices. A commercial Network Key can never be sent unencrypted through the air, whereas in the residential mode this is allowed. To securely transport between the Trust Center and a device, the Trust Center encrypts the Network Key using a Link Key based on the AES algorithm and sends the encrypted data. The device decrypts the sent data by using the Link Key. The Link Key generated for the two devices is identical for both encrypting and decrypting a message.

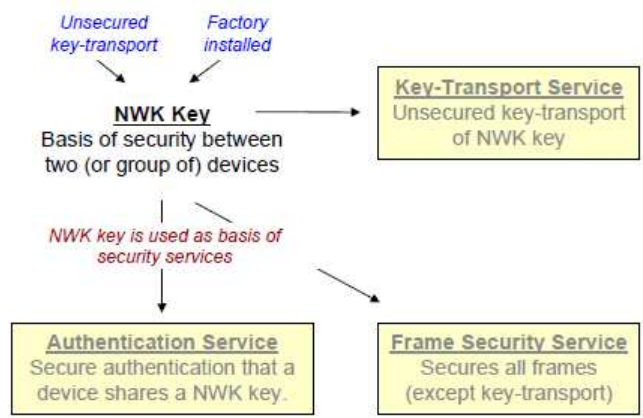

**Figure 4:** ZigBee Residential Mode[34]

ZigBee is used in the kinetic world more than any other wireless communication technology. Manipulating the physical world through wireless networks will introduce new risks as well as repeat many past mistakes [31]. There have been several vulnerabilities reported on ZigBee modules. One such vulnerability was caused by the pseudo-random number generation used in encryption. This is a real concern because this weakness will allow attackers to eavesdrop on wireless communications and potentially access the devices. The numbers generated by the random number generator (PRNG) used for asymmetric encryption were predictable. Uncovering the calculated elliptic curve cryptography (ECC) key is easier and with this information, the attacker can possibly crack the AES key for symmetric communication with other ZigBee modules since the data is transferred using ECC encryption. The problem with the protocol is a 16-bit seed used to initialize the PRNG is too short and this seed, even though derived from the converted radio module analog signal, generates values that are not scattered. Moreover, some ZigBee chips applied an insecure version of the linear feedback shift register (LFSR) to generate the pseudo random numbers. As long as the module is turned on, the LFSR always generates random numbers from the same seed and does not reseed. One of these chips containing the insecure properties produced by Texas Instruments, CC2430, has this warning on the datasheet. "This product is not recommended for new designs. However, the device, tool, or software continues to be in production to support existing customers. TI does not recommend using this part in a new design. TI encourages designers to consider alternative products for new designs" [28]. Any existing application harboring these products is vulnerable. Other researchers have also demonstrated that combined attacks are effective on supposedly secure AES implementations [5].

There are many caveats present in ZigBee networks. The mesh network model is physically accessible to many external devices and its environment is not predictable. Each node may also have many applications executing simultaneously using the same supposedly trustworthy communication transceiver. During the initial phase of the addition of a not configured device to the network, keys can be transmitted in plaintext depending on the mode.

Other researchers have identified other critical security issues concerning the ZigBee specification. When a node leaves the network (for maintenance, or being lost or compromised), the node still is able to access the network due to onboard keying information that was not properly revoked. Exploitation of the node's stored keys is a major concern. Identified as an open and interoperable model, ZigBee allows manufacturers, distributors and even end users to issue certificates (CBKE scheme). Consequently, a meter must be equipped with certificates for all potential subject certificates. This requirement raises a scalability problem due to limited storage resources of the end devices. Each device should store the root key of every possible certification subject to be able to authenticate. Storing certificates is reasonable with a coordinator node but not for the end device that has scarce storage resources [9].

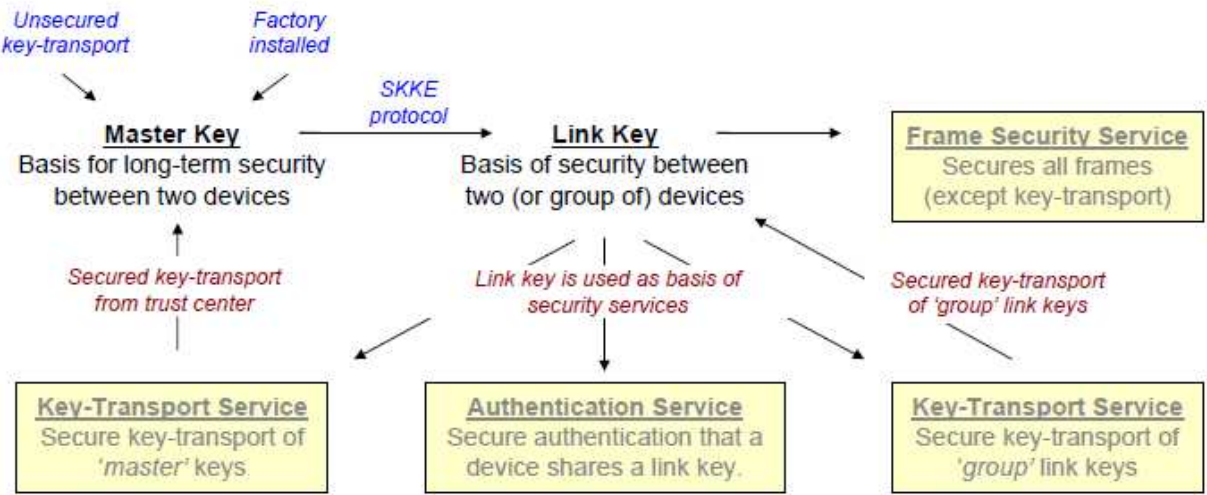

**Figure 5:** ZigBee Commercial Mode[34]

**Communication with the Collector**

As we have already discussed, the three basic logical components composing the network are the meter, the repeater, and collector. The repeater can be a specialized meter.. An intermediate data concentrator or repeater meter can be present between the smart meter and the collector. More specifically, a collector as the name implies is a device which can collect data and transmit this data in bulk at a later time. A collector has not metering functions. Meters do not have enough memory to store data. Therefore they can act as repeaters but not as collectors. A local network of repeater meters can be thus established between meters for the purpose of forwarding usage information to a collector. Collectors are typically deployed on light poles where they can be easily powered and installed. The number of collectors and repeaters is heavily dependent on environmental factors, such as the natural or building terrain, lighting, or foliage. The collector component of an AMI system performs periodic collection and consolidation of metered data from the connected end-devices based upon the configuration of their collection parameters. The collector also operates as a gateway been various WAN backhaul networks. Most deployments utilize a mesh network to transport data from the meters to the repeater (or repeater network) or the connector. Wireless mesh networks often consist of mesh clients, mesh routers and gateways. In the ZigBee specification, a node can be a coordinator, router or an end device, fulfilling the functions of the specialized nodes necessary to construct the collection network.

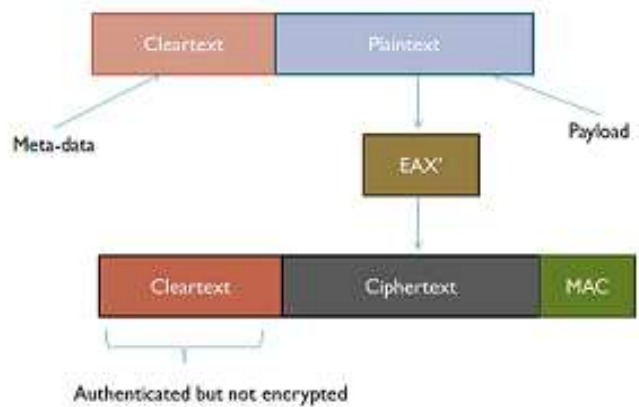
**Figure 6:** EAX' Encryption

Security protocols for communication between the smart meter and the collector are defined by ANSI standards. According to the ANSI C12.22 standard, the smart meter and the collector use a symmetric-key encryption protocol for securing their communication based on a variant of the Advanced Encryption Standard (AES) applying a modification of the EAX mode labeled EAX' for the transport of meter-based data over a network as shown in Figure 6. This standard can support either session-less or session-based connections. With symmetric encryption, authenticity is achieved through CMAC (Cipher-based Message Authentication Code) and integrity is achieved through HMAC (Hash-based Message Authentication Code) within an authenticated session. As shown in Figure 6, EAX' uses cleartext, a part of the ANSI C12.22 message that is authenticated but not encrypted, and plaintext, the message encrypted for security, to produce the Message Authentication Code (MAC).Researchers have proposed a block-cipher mode of operation for EAX' that optimizes protection in small embedded automation devices. Such optimization is crucial due to the variation of extremely large and extremely small messages [16].

The optical port on the smart meter follows the ANSI C12.18 protocol specification. An attacker with access to the firmware of one meter could extract its password and this password is likely to be shared among meters, thereby giving the attacker the ability to communicate with other meters over infrared. Also every node contains a table of symmetric keys that also could have been extracted and now this table can be used to access other meters.

Each smart meter talks to the collector and the collector in turn will send the information to the utility company. The collector can protect the messages sent to the utility company with a symmetric encryption protocol, the General Packet Radio Service (GPRS) protocol, a standard for wireless communications. The GPRS protocol aims at protecting the network against unauthorized use and the privacy of the users. The goal of the GPRS service provider is to ensure that the subscriber is the real GPRS subscriber and the subscriber wishes to access the service without having its privacy compromised, that is, with data confidentiality assurance. AGPRS session uses a symmetric key-based protocol that supports three different cipher algorithms, that is, the A3, A5, and A8, as shown in Figure 7. The necessary algorithms and keys for authentication and encryption are stored in the subscriber identity module (SIM) of the concentrator and the Authentication Center (AuC) of the GPRS network concentrator. A smart card is used to keep the subscriber's master key secret, a 128-bit key, which is protected by 4-digit pin. The network authenticates the SIM ($Mk_1$) to protect against cloning. The SIM and AuC contain both the A3/A8 algorithms to generate the related keys for authentication and ciphering respectively. The algorithm A5 ciphers the data. The ciphering is performed by the Logical Link Control protocol which is transported between the meter and GPRS support node.

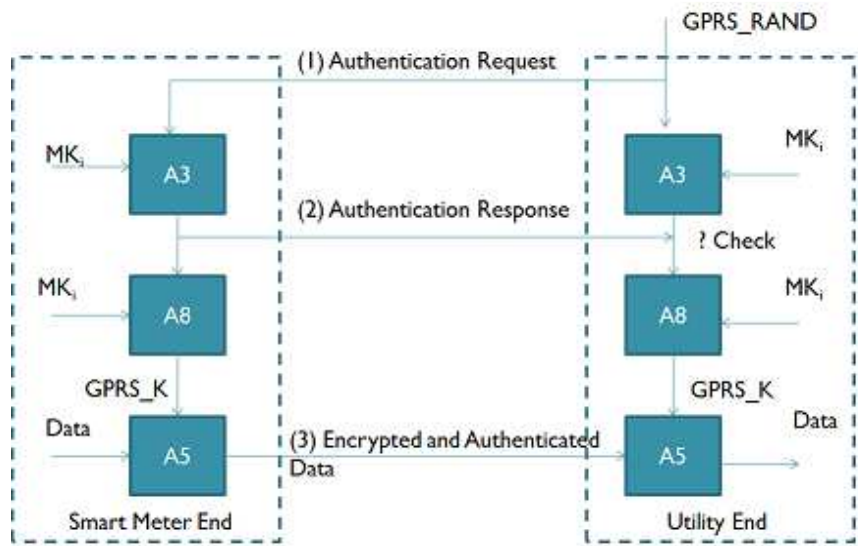

**Figure 7:** A GPRS Protocol Session

Authentication in GPRS refers to checking the identity of the meter before it can use the network resources. The GPRS operator wants to verify that the user possesses a valid SIM card with a valid subscriber key. This process should be completed without sending the SIM card key over the radio interface. As the session proceeds, the GPRS support node, located at the Utility end, sends a message to the smart meter end, usually the concentrator, containing the International Mobile Subscriber Identity (IMSI) number of the subscriber and requests for three components of the communication. The IMSI serves as a fixed subscriber number to identify the subscriber towards the network. IMSI is also stored in a database that contains packet domain subscription data and routing information and the AuC. The three components are: the RAND, a randomly generated 128 bit number to ensure that the set of components are always different; the SRES, a 32-bit number generated by the A3 algorithm and used as digital signature of the meter; and the GPRS key, a 64-bit ciphering key generated by the A8 algorithm used for encrypting data between the meter and the GPRS support node.

After receiving the components from the AuC, the GPRS support node sends the RAND number to the smart meter end for authentication. The SIM generates another SRES using the A3 algorithm, this one based on the RAND and subscriber key. The Smart meter end transmits its SRES value to the GPRS support node. The GPRS support node compares the received SRES with the AuC SRES. If both values agree, the authentication is successful. Each execution of the A3 algorithm is performed with a new value of the RAND. Since the RAND cannot be predetermined, this results in stopping the recording of channel transmission in attempts to fake an identity by playing it back.

The messages transmitted before the authentication is completed are unencrypted and this is a security concern. The process for validating authentication credentials between a meter and NAN requires many steps and these often take place before the use of encryption. The adversary can manipulate the authentication process to impersonate legitimate devices and expose the network to denial of service conditions or gain information such as cryptographic information to decrypt messages. If an attacker manages to intercept the components as they are transmitted and with knowledge of A3, it is able to reverse the calculation to derive the subscriber's key. Additionally, certain implementations of the A3 and A8 algorithms are known to be weak and extraction of the subscriber's key was completed within one hour [8].The GPRS protocol only performs one-way authentication. It does not assure that the collector is connected to an authentic utility company. Also SIM cards can be stolen. In spring 2011 a women stole the 3G SIM card from an electric meter and plugged it into her laptop. She accumulated $193,187 worth of data fees and is now serving six months in jail [4].Episodes such as these, point to fact that physical identifications are not more secure and may be easier to abuse and confiscate.

The information the smart meter sends to the collector is private and can cause security concerns if it is leaked. The encryption also has an option to be based on a session created for each communication with the communicator. The communication is through a broadcast medium. The range of this communication is not always large enough for all the smart meters to reach a communicator. Therefore, to address this issue, smart meters must be able to forward messages to and from other smart meters. This requirement creates concerns because of smart readers having access to other smart meter data.

## 6. Security Threat Model

It is essential to understand security threats and vulnerabilities in the AMI before investigating key security solutions including key management systems. A threat is anything purposeful or accidental that can disrupt the working of a network or system. A vulnerability is a weakness in the design, configuration, implementation or management. Vulnerabilities make networks or systems susceptible to threats. Finally an attack is a specific action used to exploit an existing vulnerability [25].Many attacks share common threat vectors that allow them to achieve their ends as seen in Table 3. Mapping the attacks to a threat allows for better placement of defensive mechanisms. Adversaries can either instigate passive or active attacks as shown in Table 3. In passive attacks, the goal is to obtain worthwhile information without disturbing the network to perhaps use the information in a subsequent attack. Attacks can also be classified according to the core principle that is violated: confidentiality, integrity, availability, authentication, and non-repudiation [6][15].

**Table 3: Security Threats, Attack (types) and possible countermeasures**

| Threat | Attack | Attack Type | Possible Countermeasures |
|---|---|---|---|
| Eavesdropping | Traffic Snooping | Passive– confidentiality | Strong authorization. Strong encryption. Secure comm. links with protocols that provide message confidentiality. All secrets (credentials) encrypted |
| Traffic Analysis | Traffic Analysis | Passive– confidentiality | (same as eavesdropping) |
| Modification | Altering Radio Route Table Man-in-the-Middle | Active- integrity | Data hashing and signing. Digital signatures. Strong authorization. Tamper-resistant protocols. Secure comm. links with protocols that provide message integrity. |
| Masquerade | Spoofing Sybil | Active– authentication | Strong authentication. All secrets (credentials) encrypted. |
| Replay | Rushing Short Circuit | Active– authentication | Re-authenticate. Expire sessions appropriately. |

| Denial of Service | Denial of Service | Active– availability | Use resource and bandwidth throttling techniques. Validate and filter input. |
|---|---|---|---|
| Repudiation | Circumventing the logging of security events. Tampering with the security log to conceal identity | Active - Non- repudiation services | Secure audit trails. Digital signatures. |
| Destruction/Tampering | Injection | Active– integrity | Digital signatures. Strong authorization. Data hashing and signing. |
| Planted in System | Malicious code | Active– confidentiality | Update patches. Harden configurations. Whitelisting Code signing |

Two passive threats relevant for the AMI are eavesdropping and traffic analysis. In snooping, the attacker reads exposed information to gain insight into a node or network's behavior. This unprotected data can disclose node information, such as location and power usage, thereby divulging network topology. Node identification can assist in future attacks. Traffic analysis does not look at the data within a packet, but at the specific flows being broadcasted and their associated characteristics. Therefore, traffic analysis can be conducted on fully encrypted networks and critical routing nodes can be identified.

An unauthorized change or modification is an important action to guard against. Since the confidentiality and integrity of information being transmitted through an AMI system is vital to achieve the goals of an AMI, it is crucial to ensure that the original nature of information is tamper-proof [2]. There are three ways to tamper with utility data: a) while it is being recorded (via electromechanical tampering); b) while it is stored in the meter; and c) as it is in transmission across the network. A modification is an active threat. As the name implies, it results from an attacker changing information. An example is the man-in-the-middle attack. The attacker captures messages from the sender and sends modified messages to the recipient. In the AMI there are many potential entry points, physically unprotected, that are susceptible to man-in-the-middle attacks [30].

The masquerade attack can be accomplished by a spoofing or Sybil attack. The attacker adopts the identity of another trustworthy network node by modifying the packet address information to obtain confidential data. In a Sybil attack, the attacker assumes several false identities. These pretended nodes participate in different paths increasing the likelihood of obtaining confidential data.

The replay attack can be accomplished by intercepting data packets from the meters and resending (replaying) them to the connector. If the packet contains secret information and the hacker intercepts this message and resends the packets, the hacker will garner the same the privileges that the secret offered. AMIs are susceptible to replay attacks [11][14].

Denial of service attacks weakens data and service availability by exhausting resources or isolating nodes to avoid data transfer. The adversary can flood the network with forged, useless messages to use bandwidth and resources. This prevents the meter from acting on commands. For example, it can be assumed that collector nodes have a maximum number of sessions. In a published pen-test, commands to open consecutive TCP connections were attempted through the collector's listening port. By the tenth connection the connector node became unresponsive [13] and was not able to record and transmit usage data.

Non-repudiation ensures that receiver and sender cannot deny that a message has been sent or received. Accountability and non-repudiation in AMI systems is critical for all financial transactions, including actual meter information and responses to control commands that have financial implications. In accountability, often timeliness of responses is as important, for example, as setting peak load periods for rates [6]. Therefore, accurate timestamp information and continuous time synchronization across all AMI system components are crucial.

An active attacker will interfere with the operations of the network by tampering with nodes by broadcasting, sending messages and causing collisions. Active attacks have the potential to cause greater harm to the network as their effects may in turn affect other changes to the network. Secondly, it is important to guard against the tampering of meters. An achievable physical attack is to reverse-engineer a meter and to develop a way to modify it. Attackers can use previous cable-modem attacks that are openly available to obtain useful insights. Also tools are available that allow one to modify advanced meters, providing an easy access for attackers wishing to compromise the AMI [20]. Software tools to test (attack) the protocols can be downloaded such as KillerBee [12].

Malicious code can be inserted and executed on smart meters. Researchers also demonstrated that a computer worm could hop between meters in a power grid with smart meters, giving criminals control over those meters. It also has be verified that current AMI systems are already under hacking attacks from adversaries believed to be working overseas[7][32].

Keeping data secure will be a major challenge for both the utility company and the meter vendors. Local attackers have physical access to the meter, the network gateway and connections to these components. The attacker's intent is to disclose or alter the data in the meter or the gateway while it is being sent. Local attackers will only impact one gateway. Global or WAN attackers also may attempt to compromise meter data or the configuration data through the WAN. They may penetrate further by attempting to cause damage to a component or to the grid itself. The new levels of automation and extended access to the grid have raised issues with potential security gaps within AMI deployments. Many of the vulnerabilities AMI installations are experiencing were common in the wireless Internet. For the Internet domain, they have been somewhat mitigated [28]. Working solutions from the past must be integrated with solutions tailored for the unique characteristics of AMIs.

Clearly, as mentioned in previous sections of this report, there is a need for robust security provisions to protect AMI systems. In 2009 NIST formed the Smart Grid Interoperability Panel-Cyber Security Working Group (SGIP-CSWG). The SGIP-CSWG, whose members total more than 475 participants from the private sector (including vendors and service providers), manufacturers, various standards organizations, academia, regulatory organizations, federal agencies and international organization, provided input into *NIST's Guidelines for Smart Grid Cyber Security*[18]. From the introduction of this three volume report released in September 2010, the CSWG identified an initial set of high-priority research and development challenges arranged into four categories: device level, cryptography and key management, systems level and networking level. A common problem identified by the CSWG and other researchers is in the deficiencies in key establishment, management and distribution [9][22]. Key management must scale up to millions of credentials and keys for the sensors, meters and other AMI equipment and to a broad range of technologies used to implement the AMI.

Thinking of security as a chain, it is only as strong as its weakest link. The links are security protocols, cryptography, and key storage. Many security protocols used in AMI have been used and tested on other systems. When correctly implemented, they fulfill their intended goal. Cryptography is seldom the weakest part of a software system. Even if a system employees a weak system, an attacker can probably find much easier ways to break the system than attacking the crypto. Even though systems are definitely breakable through a concentrated crypto attack, successfully carrying out the attack usually requires a large computational effort and some knowledge of cryptography. Any encryption algorithm is only as secure as its keys. As implied, data encryption can only be part of the solution. The end user must still deal with distributing, securing, and renewing decryption keys. Thus, encryption simply shifts the burden from protecting data to protecting keys [1].

The cornerstones of the security architecture are the keys. As such, their protection is of paramount importance and keys are never supposed to be transported through an insecure channel. The basic mechanism to ensure confidentiality is the adequate protection of all keying material. Trust must be assumed in the initial installation of the keys as well as in the processing of security information.

## 7. Initial Research Directions

The previous discussions clearly indicate that application-level encryption is crucial to protect the confidentiality and integrity messages transmitted between smart meters and the utility. As shown in Figure 7, the messages between smart meters and the utility are transmitted across multiple networks. These messages go through one or more collectors and possibly through other smart meters which act as routing nodes. Different network segments use different communication protocols and they have their own transport level security. For example, the communication link between collectors and the utility may use GPRS which provides its own transport level symmetric key encryption and authentication protocols. When messages go from one network to another, even if each network provides its own network level security, the payload of the message is still exposed to the routing node as it removes the previous security layer and adds a new security layer. Further, certain network security protocols, such as GPRS, are known to be weak and can easily be compromised [38]. If messages travel through such networks, the payloads of the messages may be exposed to adversaries. If AMI systems provide application level end-to-end security, actual data are still protected when such exposures of payloads occur.

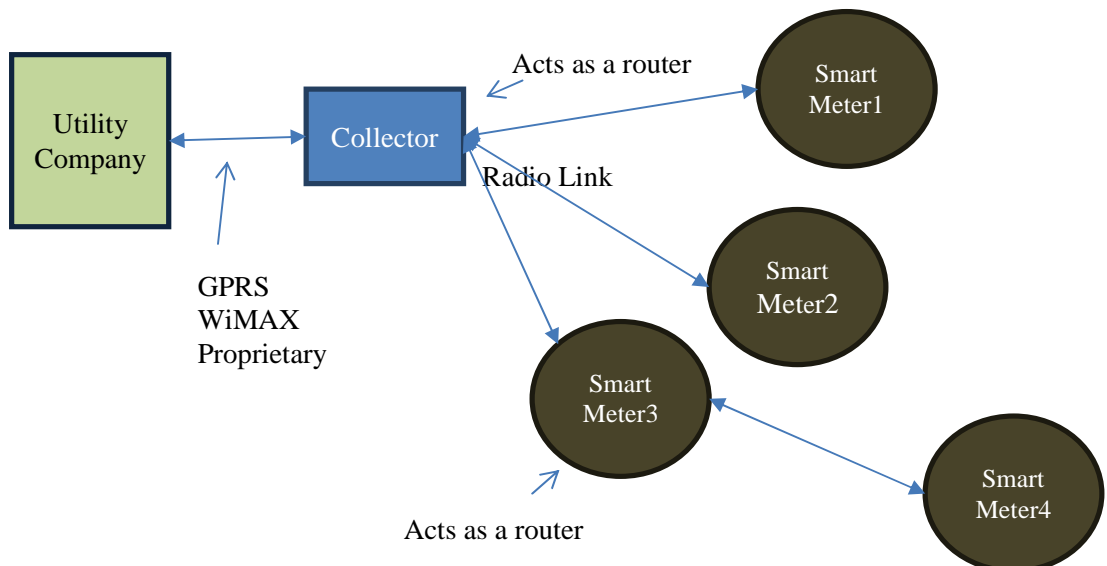

**Figure 7:** A Simplified AMI System

A scalable and efficient key management scheme is thus essential to provide application level security in AMI systems. As AMI systems potentially consist of thousands of smart meters, it is thus important that, in addition to protocols like the ANSI C12.22, scalable group key management schemes be supported. These schemes are typically used for broadcast communications. In addition, mechanisms must be provided for authenticating smart meters and generating secrets. In what follows, we briefly describe initial approaches to the above mentioned issues.

**Group Key Management Schemes**
The scalability of the group key management (GKM) scheme used is important to support the operations of such large systems. The scalability mainly depends by the following factors of the GKM scheme:

- The size of the ciphertext.
- The number of secrets needs to be stored at smart meters, the collectors, and the utility.
- The efficiency of the GKM operations, especially re-keying.

In order to prevent network bottlenecks, the size of the ciphertext should be sub-linear or at least linear in the size of the AMI system. Since smart meters have limited storage capacity, it is important that the GKM scheme supports minimal key storage at smart meters. The efficiency of the re-keying operation decides how many or what fraction of smart meters are affected by that operation. With a large group of smart meters, re-keying operations should be efficient to minimize the overhead on the AMI system. All the GKM operations, especially the smart meter side operations, should be computationally efficient, as smart meters have limited processing capabilities.

Trivial GKM schemes would not work in such large AMI systems. One approach is to use a public key infrastructure (PKI) based solution where each smart meter has a unique public/private key pair. Another approach is to assign each smart meter a unique private key and use symmetric key encryption. Yet another approach is to use a single symmetric key and share it with all smart meters. These approaches suffer from one or two weaknesses: inefficient client side GKM operations, large ciphertexts and inefficient re-keying. Therefore, we need a different solution.

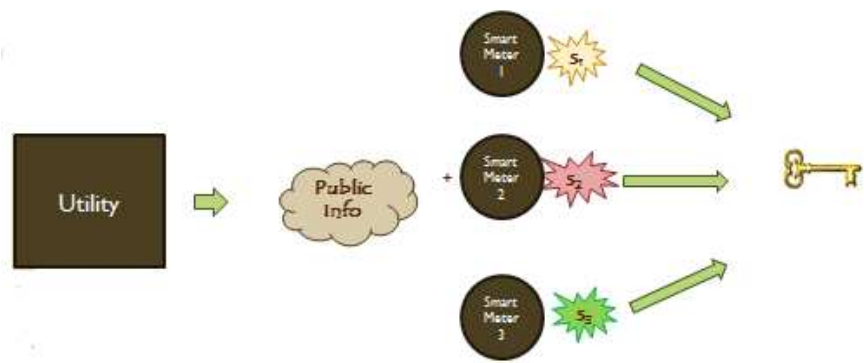

**Figure 8:** A High-Level View of BGKM Key Derivation

We propose to use a broadcast GKM (BGKM) due its desirable proprieties to efficiently handle large AMI systems. In such a scheme, the utility acts as the group controller and smart meters as clients. As shown in Figure 8, with BGKM schemes, smart meters are not given symmetric keys instead they are given secrets. The utility broadcasts some public information, which is constructed from those secrets, to the AMI system, and smart meters having valid secrets can derive the symmetric key using their secrets and the public information. Most BGKM schemes [35, 36, 37] demonstrate the following desirable properties:
- Ciphertext is linear or sub-linear in the group size of the targeted smart meters.
- Designed for symmetric key cryptography.
- Re-keying does not affect other smart meters.
- Each smart meter needs to store one or a few secrets only.
- Client side GKM operations are efficient.

A BGKM scheme consists of the following five algorithms:
- Setup – it initializes system and security parameters.
- SecGen – it assigns one or more secrets to each client.
- KeyGen – it generates the public information using the secrets of a set of selected clients. The public information hides the group key.
- KeyDer – A client with a valid secret uses this algorithm to derive the group key from the public information.
- Re-Key – it re-generates the public information using the updated set of secrets. The new public information hides a new group key.

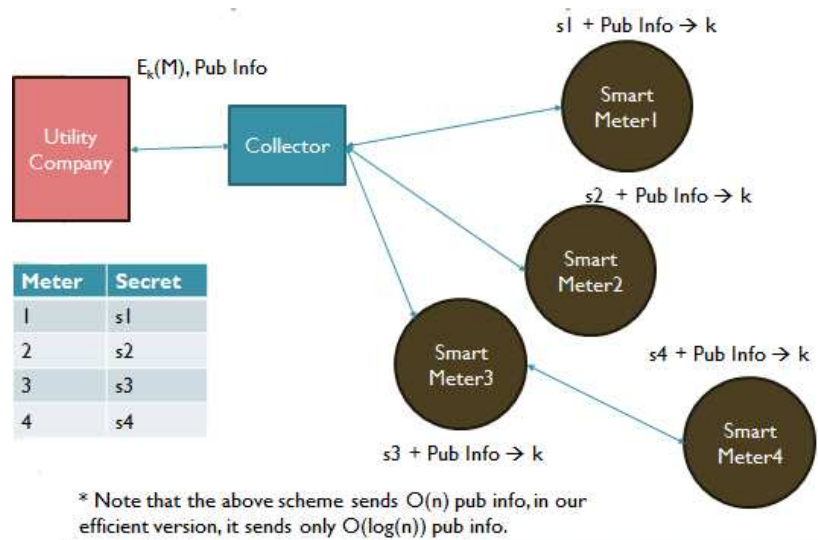

**Figure 9:** Using BGKM with an AMI System

Figure 9 shows a simplified view of using a BGKM scheme with an AMI system. Initially, the utility executes the Setup algorithm to initialize the system parameters. As shown in the table, the utility gives each of the four smart meters a secret using the SecGen algorithm. These secrets must be unique in order to support other BGKM algorithms. When the utility wants to broadcast a message to a selected set of smart meters (to all smart meters in this example), it selects the corresponding secrets and executes the KeyGen algorithm. It produces a symmetric key k and public information "Pub Info". It broadcasts the "Pub Info" and the encrypted message $E_k(M)$ to the AMI system. Any of the smart meters in that group can derive the key using the KeyDer algorithm giving their secret and the "Pub Info" as the input. Once the key is derived, they can decrypt the message. Notice that the collector cannot derive the key and decrypt the message, as it does not possess a valid secret. When a smart meter is revoked, the utility simply needs to remove the corresponding entry from the table storing the secrets and this operation does not affect other smart meters. When a new smart meter is added, the utility simply needs to add a secret to the table and this operation also does not affect other smart meters. After adding or revoking one or more smart meters, if the utility should provide backward secrecy or forward secrecy, it has to invoke the Re-Key algorithm before sending any new messages.

**Authenticating Smart Meters and Generating Secrets**
The BGKM based solution protects the confidentiality and the integrity of the messages transmitted over the AMI network. However, the security of this solution depends on one important security function: the ability of the utility to correctly authenticate smart meters. The conventional approach to authenticate a device is to place a secret key in non-volatile memory inside the device and use cryptographic primitives such as digital signature. Unfortunately, such an approach suffers from a few drawbacks. An adversary may physically extract secret keys from non-volatile memory. Further, these cryptographic operations are too costly for resource constrained devices like smart meters. Therefore, AMI networks require a different approach to authenticate smart meters.

We propose to integrate a PUF (Physically Unclonable Function) device with each smart meter in order to implement a hardware-based, low cost and secure authentication mechanism. PUF devices derive secrets from complex characteristics of integrated circuits instead of storing the secrets in volatile memory. As PUF relies on the random variation during the integrated circuit fabrication process, even two PUFs with the same layout result in two different secrets. Since PUF produces volatile secrets, it is extremely difficult to carry out invasive attacks such as extracting the key. Further, different PUF devices produce unique secrets, it is very difficult to clone a PUF and launch spoofing attacks. PUF can be considered as a function that maps a set of challenges to a set of responses based on an intractably complex physical system. PUF demonstrates the following characteristics that allow one to use it as an authentication mechanism:

- Different challenges to the same PUF produce unique responses.
- Same challenge to different PUFs produces unique responses.

Our approach to perform the PUF based authentication is as follows. During the time of installation of smart meters, the utility stores a challenge-response pair for each smart meter. The utility authenticates a smart meter based on its ability to produce the correct response for the stored challenge. Due to the aforementioned characteristics, it is difficult to launch spoofing attacks.

We use the same PUF mechanism to generate secrets required for the BGKM scheme. The idea is to add a feedback loop to the PUF to feed the response as a new challenge and run the loop for a selected number of iterations. Due to the characteristics of PUF mentioned earlier, it produces a unique secret with a high probability. This secret generation mechanism has the benefit of not having to store secrets in the meter, but instead generate them as and when needed. Therefore, the secrets are protected from the invasive attacks.

## 8. Conclusions

Due to various benefits of AMIs, smart meters have been increasingly deployed. However, security issues in transmitting data between the utility and smart meters have been a major concern. In this work, we surveyed the security of smart meters and the current communication standards and protocols used, and then proposed a BGKM and PUF based approach to correctly authenticate smart meters and support end-to-end confidentiality and integrity of messages transmitted between the utility and smart meters through a public AMI network.